\documentclass{article}
\usepackage{spconf,amsmath,graphicx}

\usepackage{enumitem}
\setlist{nosep, leftmargin=14pt}

\usepackage{mwe} 
\usepackage{graphicx}
\usepackage{comment}
\usepackage{multicol}
\usepackage{multirow}
\usepackage{booktabs}

\title{Relation U-Net}
\name{Sheng He, Rina Bao, P. Ellen Grant, Yangming Ou
}
\address{Boston Children's Hospital, Harvard Medical School}

\begin{document}
%
\maketitle
\begin{abstract}
Towards clinical interpretations, this paper presents a new “output-with-confidence” segmentation neural network with multiple
input images and multiple output segmentation maps and their
pairwise relations. 
A confidence score of the test image without
ground-truth can be estimated from the difference among the
estimated relation maps. We evaluate the method based on the
widely used vanilla U-Net for segmentation and our new model
is named Relation U-Net which can output segmentation maps
of the input images as well as an estimated confidence score
of the test image without ground-truth. Experimental results on
four public datasets show that Relation U-Net can not only provide better accuracy than vanilla U-Net but also estimate a confidence score which is linearly correlated to the segmentation accuracy on test images.
\end{abstract}
\section{Introduction}
\label{sec:intro}

Many deep learning-based methods for medical image segmentation~\cite{ronneberger2015u}
have been proposed to improve the accuracy. 
Most deep learning models are
“output-only” models, which only output a single segmentation map given the input image. 
In general, the average accuracy on the evaluation dataset is reported.
However, in practice the average accuracy on the evaluation
dataset does not mean that every test image has the same
or similar accuracy, especially the test images sampled from
the clinical practice. Some test images are easy samples with
high quality while some test images are difficult samples
with smooth boundaries, poor contrast or small size of the
target regions with high uncertainty~\cite{mehrtash2020confidence}. 
Thus, deploying
the trained model in clinical practice requires an interactive
interpretation or post-correction of the segmentation outputs to
automatically identify a subset of the difficult testing samples
for experts inspection~\cite{agarwal2022estimating}.
There is an unmet need to evaluate segmentation
accuracies and even to reject failed segmentation when, in
real-world applications, the ground-truth is often unavailable.

To address this problem, an “output-with-confidence” model
is required in real-world applications.
It needs to automatically
rank the test samples without ground-truth by the difficulty or
the confidence score of segmentation on each sample image.
Based on the rank list, end users can select the more challenging testing samples for further inspection~\cite{agarwal2022estimating}, which can save the
human annotation and auditing time. 

\begin{figure}[!t]
    \centering
    \includegraphics[width=0.5\textwidth]{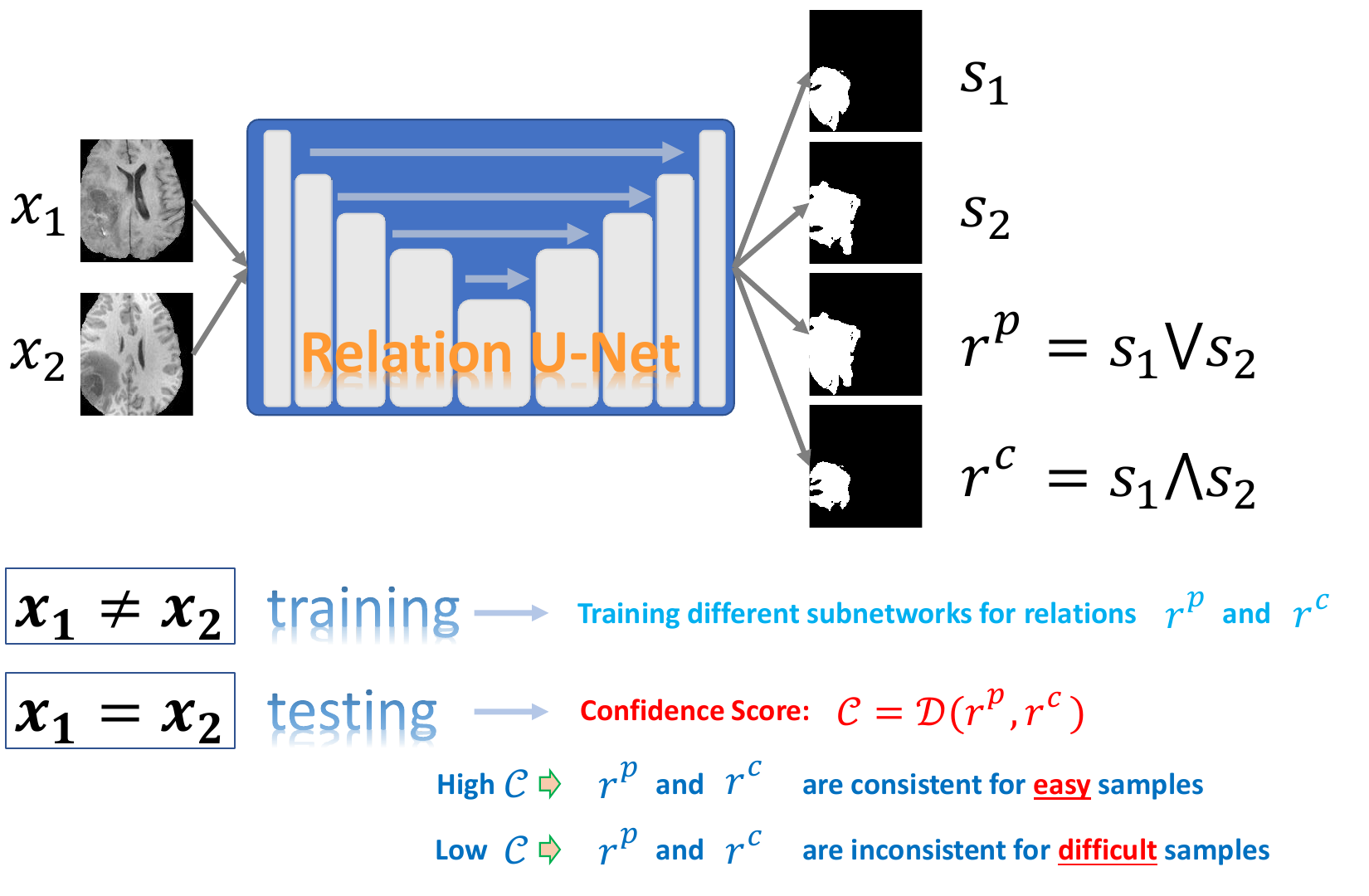}
    \caption{Illustration of the proposed Relation U-net.}
    \label{fig:runet}
\end{figure}

\begin{figure*}[!t]
    \centering
    \includegraphics[width=\textwidth]{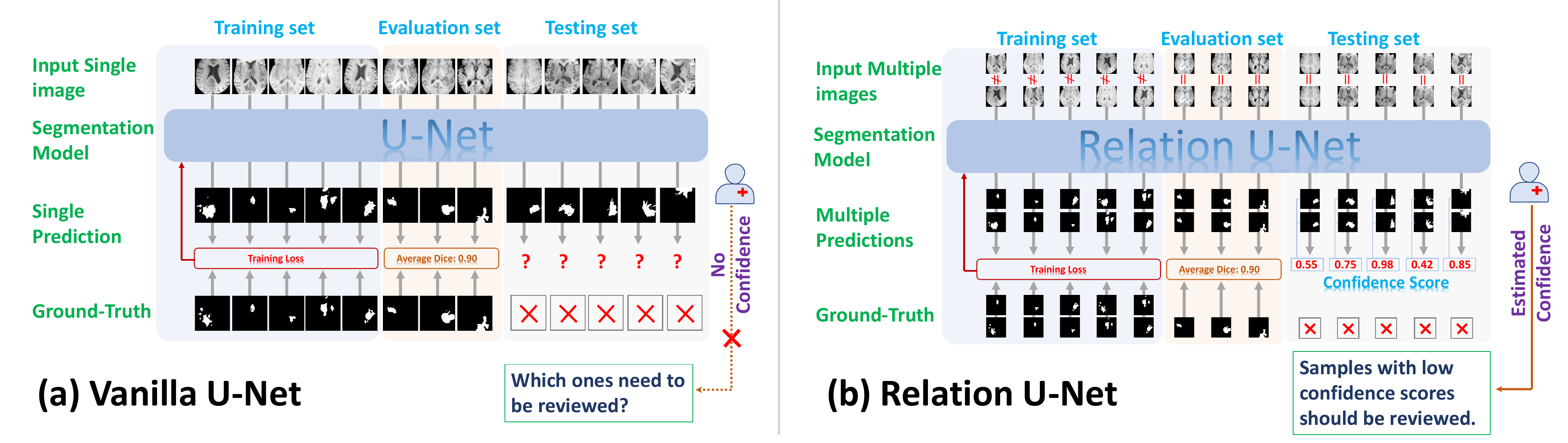}
    \caption{The comparison of the vanilla U-Net (a) and the proposed Relation U-Net (b). (Zoom in for a better visualization)}
    \label{fig:framework}
\end{figure*}

In this paper, we propose an example of the “output-with-confidence” model based on the popular U-Net for segmentation, named Relation U-Net (Fig.~\ref{fig:runet}).
It is a generalization of the vanilla
U-Net and Fig.~\ref{fig:framework} shows the main difference between them. The proposed Relation U-Net receives two input images and outputs
four segmentation maps including the segmentation of
each input image and their corresponding relations.
During testing, the two input
images can be sampled from the same input image and
the discrepancy of the estimated
relation segmentation maps is correlated to the segmentation
accuracy of each test images. Thus, the discrepancy of the
estimated relation segmentation maps can be considered as a
confidence score which can be used to
rank the test images for the end users to inspect and review
the difficult samples.

\section{Method}

\subsection{Relation U-Net}

Fig.~\ref{fig:runet} shows the framework of the Relation U-Net.
Given two different input images $x_i$, $i=1,2$ their
corresponding binarized segmentation maps $s_i$
in the training set, we use logical operations to define the relations based
on segmentation maps $s_i$
, similar to studies~\cite{wang2021pairwise}. Two
relations are considered in this paper, including the \textit{Possible}
(union) $r^p=s_1\vee s_2$ and \textit{Consensus} (intersection) $r^c=s_1\wedge s_2$ relations between
the segmentation maps $s_1$ and $s_2$ of two different images $x_1$
and $x_2$.
The \textit{Possible} relation map captures the
pixels/voxels that have been segmented as belonging to the
target in at least one image. The \textit{Consensus} relation map
represents the pixels/voxels that have been segmented as
belonging to the target object in both images.

\subsection{The structure of neural network}

In this paper, we use the multiple-input multiple-output
(MIMO) configuration~\cite{wang2021pairwise,he2022deep} to build Relation U-Net. We use the
popular U-Net~\cite{ronneberger2015u} as the baseline which only requires two changes. At the input layer, the two input images
are concatenated. Since we need four outputs (see Fig.~\ref{fig:framework}), at the last layer, four different predictions are
obtained by adding four different classifiers. 

The Relation U-Net takes two inputs and returns four outputs
including the estimated segmentation $s_i$ for the input images
and their corresponding relations $r^p$ and $r^c$. 
During training,
the two inputs are sampled independently from the training set
and the four different classifiers for segmentation are trained
based on the ground-truth of corresponding labels. Thus, the
Relation U-Net does not only learn the information for each
input image, but also learns the relations among the pairs
of samples. At test time, the multiple input images can be
sampled from different images or the same input image is
repeated two times to feed into the Relation U-Net, which can
make four independent segmentation maps.

\subsection{Estimating of Confidence Score}

\begin{figure}[!t]
    \centering
    \includegraphics[width=0.5\textwidth]{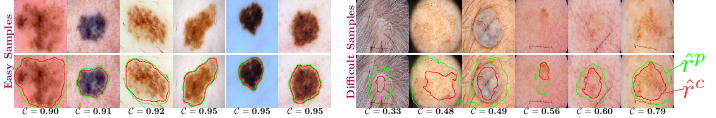}
    \caption{Examples of images on the ISIC2018 dataset with relation segmentation maps: $\hat{r}^p$  (green contours) and $\hat{r}^c$ 
(red contours). The $\hat{r}^p$ and $\hat{r}^c$ 
are the same on easy samples while are different on difficult samples. $\mathcal{C}$ measures how the consistency of the segmentation maps between $\hat{r}^p$ and $\hat{r}^c$.}
    \label{fig:isicexa}
\end{figure}

Based on the definition , the \textit{Possible} $r^p$ and
\textit{Consensus} $r^c$ relations are reflexive~\cite{wang2021pairwise}, indicating that
when the pair of input images $x_1$, $x_2$ are from the same sample $x_1=x_2$, the output of the relations $r^p$ and $r^c$
is the same to the segmentation of the input: $r^p=s^c=s_1=s_2$.
However, since we train the Relation U-Net with two different input images independently sampled from the training
set and the four outputs are predicted independently.
Thus, the
estimations of the \textit{Possible} $\hat{r}^p$
and \textit{Consensus} $\hat{r}^c$
are not exactly the same given the same input since they are independently
predicted by different sub-networks of the Relation U-Net. We
compute the discrepancy between the estimation of $\hat{r}^p$ and $\hat{r}^c$
by: $\mathcal{C}=\mathcal{D}\big(\hat{r}^p,\hat{r}^c\big)$.
where $\mathcal{D}$ is a distance metric to measure the difference between
two estimated segmentation maps of $\hat{r}^p$ and $\hat{r}^c$.
In this
paper, we use the Dice coefficient as the distance metric to
measure the difficulty to segment each image, defined as:
$\mathcal{C}=2|\hat{r}^p\cap\hat{r}^c|/(\hat{r}^p+\hat{r}^c\big)$.
where $\mathcal{C}$ measures the consistency between the segmentation
maps between $\hat{r}^p$ and $\hat{r}^c$.
Fig.~\ref{fig:isicexa} shows several images with
the segmentation maps of the relations $\hat{r}^p$ and $\hat{r}^c$.
For difficult
samples, the segmentation maps of these two relations are
different because they are independently predicted by different
sub-networks within the Relation U-Net, yielding a low $\mathcal{C}$.
However, for easy samples, the segmentation maps of these
two relations are almost same and the corresponding $\mathcal{C}$ values
are also high. Thus, the value of $\mathcal{C}$ can be used to measure the
difficulty of the test images which is correlated to the accuracy
of the segmentation result. In this paper, we name the value
of $\mathcal{C}$ as the confidence score which can rank test images by difficulty without ground-truth.

\section{Experimental results}

\subsection{Datasets}

We use four public datasets to evaluate the proposed method, including LiTS~\cite{bilic2023liver}, Hippocampus~\cite{antonelli2022medical}, BraTS~\cite{menze2014multimodal} and ISIC~\cite{codella2019skin}.
The LiTS dataset contains contrastenhanced CT images.  The Hippocampus dataset contains
MRI images with the anterior and posterior of the hippocampus. 
For BraTS, we use the brain MRIs
on 2018 Brain Tumor Segmentation (BraTS) challenges~\cite{menze2014multimodal}
with four different modalities (T1, T1ce, T2 and FLAIR). The
ISIC contains dermoscopic images with skin cancer lesions.
We use five-fold cross-validation strategy: all datasets
are randomly split into five folds. Each time, one fold is used
for testing and the rest of four folds are used for training.
Average accuracy and standard deviation are reported.

\subsection{Implementation details}

We follow the common strategy to train the neural network
without loss of generality. Specifically, the Relation U-Net is
trained by the Adam optimizer with the cross-entropy loss.
The initial learning rate is set to 0.0001, and reduces to half
at every 20 epochs in the total 100 training epochs. 
For Relation U-Net, we randomly sample two images on the training set during training. 
The ground-truth of the Possible and Consensus is computed based
on the ground-truth of the input images for training. For evaluation, we
use the common Dice Similarity Coefficient (DSC)~\cite{mehrtash2020confidence,wang2021pairwise}.

\subsection{Accuracy of Relation U-Net}
\begin{table}[!t]
    \centering
    \caption{The Dice performance of U-Net and Relation U-Net on the four datasets when $x_1=x_2$.}
    \label{tab:diceres}
    \resizebox{0.5\textwidth}{!}{
    \begin{tabular}{l|c|c|c|c|c|c}
    \toprule
       \multirow{2}{*}{Dataset} &  \multirow{2}{*}{U-Net~\cite{ronneberger2015u}} & \multicolumn{5}{c}{Relation U-Net} \\
       \cmidrule{3-7}
       & & $\hat{s}_1$ & $\hat{s}_2$ &($\hat{s}_1$+$\hat{s}_2$)/2 & $\hat{r}^p$ & $\hat{s}^c$ \\
       \midrule
        LiTS &  91.97{\tiny$\pm$0.69} & 92.44{\tiny$\pm$0.61} & 92.42{\tiny$\pm$0.91} & 92.62{\tiny$\pm$0.72} & \textbf{92.78}{\tiny$\pm$0.74} & 91.78{\tiny$\pm$0.83} \\
        Hippocampus &  82.76{\tiny$\pm$0.42} & 82.88{\tiny$\pm$0.76} & 82.34{\tiny$\pm$0.56} & 82.59{\tiny$\pm$0.65} & \textbf{83.24}{\tiny$\pm$0.48} & 78.60{\tiny$\pm$0.74} \\
        BraTS &  82.63{\tiny$\pm$0.97} & 82.91{\tiny$\pm$0.88} & 82.96{\tiny$\pm$0.83} &  83.27{\tiny$\pm$0.86} & \textbf{84.24}{\tiny$\pm$0.78} & 80.92{\tiny$\pm$1.16} \\
        ISIC &  87.89{\tiny$\pm$0.53} & 87.36{\tiny$\pm$0.68} & 87.47{\tiny$\pm$0.56} &  \textbf{87.93}{\tiny$\pm$0.54} & 86.08{\tiny$\pm$0.97} & 86.33{\tiny$\pm$0.84} \\
    \bottomrule
    \end{tabular}}
\end{table}

\subsubsection{Accuracy of Relation U-Net when the input images
sampled from the same image} 
Table~\ref{tab:diceres}  shows the Dice values of the vanilla U-Net (baseline) and the proposed Relation U-Net with estimated outputs of $s_1$ and $s_2$ and their relations $r^p$ and $r^c$.
$(s_1+s_2)/2$ denotes
the average of the $s_1$ and $s_2$ in the testing when the input
images $x_1$ and $x_2$ are sampled from the same input. 
Two observations can be obtained: (1) The output $r^p$ provides higher Dice scores than U-Net on LiTS, Hippocampus, and BraTS and the average of $(s_1+s_2)/2$ provides higher Dice score than U-Net on ISIC.
The results indicate that the
Relation U-Net can learn the sub-networks for estimating the
segmentation maps and their relations of different input images. (2) From the table~\ref{tab:diceres}, we can see that the accuracy of the Possible relation $r^p$
is higher than the accuracy of the Consensus relation $r^c$,
indicating that these two relations capture different information
for segmentation obtained by the different sub-networks of
the Relation U-Net. 

\begin{table}[!t]
    \centering
    \caption{Comparison of Relation U-Net with other baselines.}
    \label{tab:comps}
    \resizebox{0.5\textwidth}{!}{
    \begin{tabular}{l|c|c|c|c|c}
    \toprule
     \multirow{3}{*}{Dataset} & \multicolumn{2}{c}{U-Net~\cite{ronneberger2015u}} & \multicolumn{3}{|c}{Relation U-Net} \\
     \cmidrule{2-6}
     &  \multirow{2}{*}{Baseline} & \multirow{2}{*}{Dropout~\cite{nair2020exploring}}  & $x_1=x_2$ & $x_2 \in$ Train & $x_2 \in $ Test \\
     \cmidrule{4-6}
         & & & ($\hat{s}_1$+$\hat{s}_2$)/2 & $\hat{s}_1$ & $\hat{s}_1$\\
    \midrule 
    LiTS & 91.97{\tiny$\pm$0.69} & 92.16{\tiny$\pm$0.80} & 92.62{\tiny$\pm$0.72} & 92.58{\tiny$\pm$0.64} & \textbf{92.84}{\tiny$\pm$0.39}\\
    Hippocampus &  82.76{\tiny$\pm$0.42} & 82.63{\tiny$\pm$0.60} & 82.59{\tiny$\pm$0.65} & \textbf{83.12}{\tiny$\pm$0.62} &\textbf{83.12}{\tiny$\pm$0.63} \\
    BraTS &  82.63{\tiny$\pm$0.97} & 82.73{\tiny$\pm$1.02} & 82.27{\tiny$\pm$0.86} & \textbf{83.37}{\tiny$\pm$0.95} & 83.36{\tiny$\pm$0.92}    \\
    ISIC &  87.89{\tiny$\pm$0.53} & 87.73{\tiny$\pm$0.66} & 87.93{\tiny$\pm$0.54} & 88.00{\tiny$\pm$0.51} &  \textbf{88.01}{\tiny$\pm$0.52}  \\
    \bottomrule
    \end{tabular}}
\end{table}

\begin{figure*}[!t]
    \centering
    \includegraphics[width=\textwidth]{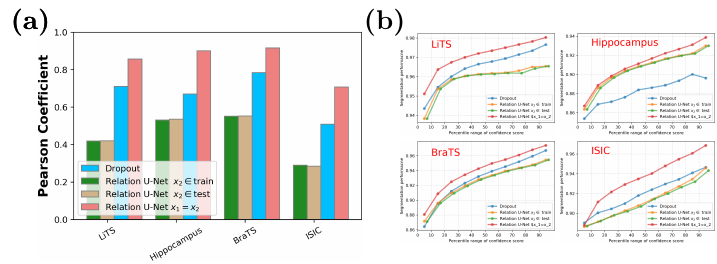}
    \caption{(a) Comparison of the Pearson coefficient between the segmentation accuracy and confidence scores. (b) The segmentation accuracy (y-axis) for the test images thresholded by the confidence score (x-axis). The results are computed by the average over the five-fold cross-validation. On the test images without ground-truth, the segmentation accuracy increases with an increase in confidence scores.}
    \label{fig:results}
\end{figure*}

\begin{figure*}[!t]
    \centering
    \includegraphics[width=\textwidth]{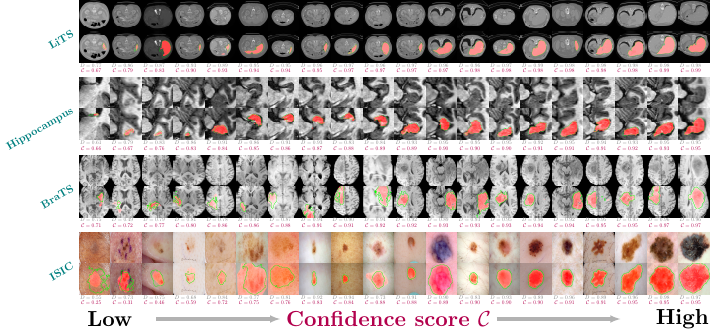}
    \caption{Examples of the images (the first row on each data set) and their corresponding segmentation results (the second row on each dataset) ranked by the confidence score $\mathcal{C}$. The red masks denote the ground-truth and the green contours denote the segmentation results. The Dice similarity coefficient $\mathcal{D}$ and confidence score $\mathcal{C}$ are shown under each example.}
    \label{fig:exmaples}
\end{figure*}

\subsubsection{Accuracy of Relation U-Net when the input images
sampled from the different images} 
In this section, we evaluate the performance of the input $x_1$ given the input image $x_2$ which is sampled from training set (denoted as $x_2\in train$) and testing set (denoted as $x_2\in test$). In this scenario, the input $x_2$ can be considered as the anchor and the different segmentation outputs $s_1$ can be obtained by varying the choice of the anchor $x_2$. For each input image $x_1$, we randomly select 200 different images as the second input $x_2$ and the average results are reported. 
We also compare it with the Dropout which trains the sub-networks during training by randomly dropping some neurons. For the vanilla U-Net, different segmentation
maps can be obtained by applying the Monte Carlo (MC) Dropout
on test images, similar to study~\cite{nair2020exploring}. We first train a U-Net with a dropout on the last layer and then take the MC samples of the prediction using the dropout during testing. We set $n=200$ and it also requires 200 forward passes for the same input image.
Table~\ref{tab:comps} shows the results of different models on the four datasets. The segmentation accuracy of the Relation U-Net
with different input images is slightly higher than the
one with the input images sampled from the same image. The
results indicate that the ensemble results of different pairs of
input images can provide a robust segmentation.  Finally, the Relation U-Net provides higher accuracies than
the Vanilla U-Net (baseline) and U-Net with dropout.

\subsection{Correlation between the accuracy and confidence score $\mathcal{C}$}

This section presents the results of the correlation between
the accuracy of the segmentation and the corresponding estimated confidence score $\mathcal{C}$. We also compare the proposed
$\mathcal{C}$ with the one computed based on the vanilla U-Net with Dropout where different segmentation maps can be obtained by applying the MC Dropout~\cite{nair2020exploring} on test images with $n=200$ forward passes.
Following study~\cite{joskowicz2018automatic}, the Possible relation of the $n$ different segmentation maps is computed as: 
$\hat{r}^p=\cup s_i$ and the
Consensus relation is computed as: $\hat{r}^c=\cap s_i$ where $\cup$ and $\cap$
are the operations of the union and intersection, respectively.

Fig.~\ref{fig:results}(a) shows the Pearson coefficients of different models which is computed between the confidence scores and the segmentation accuracies on the testing samples. It shows that the Pearson coefficient of the proposed
Relation U-net are higher than ones of other three methods
across the four datasets.
To test the discriminative of the confidence score, we plot the segmentation accuracy of test samples
bucketed by the decile of confidence score~\cite{agarwal2022estimating}. We first
rank all testing samples based on the estimated confidence
score and the testing samples within the d\% percentile are
included to compute the accuracy of the segmentation. This
is similar to the coverage~\cite{ghesu2021quantifying,he2023segmentation} which rejects the (100-d)\%
difficult samples for further attention. Fig.~\ref{fig:results}(b) shows the curves with different d\% obtained by the Relation U-Net with the input images sampled from the same image $x_1=x_2$.
We show that examples at the highest percentiles on the
rank often have high segmentation accuracy.   Fig.~\ref{fig:results}(b) shows that when the ground truth is not available, we can still use the proposed confidence score to find difficult-to-segment samples. The remaining relatively easier-to-segment cases do
enjoy a higher segmentation accuracy once the ground-truth
is available. Overall, the confidence score provides an
informative metric to evaluate segmentation accuracy when the
ground-truth is not available.

Fig.~\ref{fig:exmaples} visualizes test images and their corresponding segmentation results ranked by the confidence scores. Images with low confidence scores tend to have small target regions, smooth boundaries and poor contrast between the target and background tissues. 
The accuracies of
their segmentation results are usually low. Images with high
confidence scores often have large target regions and clear
boundaries, yielding more consistent segmentation maps of
the Possible and Consensus relations. 

\section{Conclusion}
This paper proposed a simple generalization of the segmentation neural network which can not only provide the
segmentation result, but also an estimated confidence score
to evaluate segmentation accuracies for each test image when
the ground-truth is not available. The segmentation network
receives multiple two input images and outputs multiple outputs
four different segmentation maps including the two segmentation maps of each input image as well as the two relations among the pairwise of input images. 
Our proposed method can provide better results than the baseline and can also provide a confidence score for ranking the test images by difficulty without ground-truth.

\bibliographystyle{IEEEbib}
\bibliography{finalversion}

\end{document}